\documentclass[reprint,twocolumn,aps,prl,superscriptaddress,showpacs,a4paper,amsmath,amssymb,floatfix]{revtex4-1}

\usepackage{graphicx}

\begin{document}

\preprint{}

\title{Coordination-driven magnetic-to-nonmagnetic transition in manganese doped silicon clusters}

\author{V.~Zamudio-Bayer}
 \thanks{LL and VZB contributed equally to this work.}
 \affiliation{Institut f\"{u}r Methoden und Instrumentierung der Synchrotronstrahlung, Helmholtz-Zentrum Berlin f\"{u}r Materialien und Energie GmbH, Albert-Einstein-Stra{\ss}e 15, 12489 Berlin, Germany}
 \affiliation{Institut f\"{u}r Optik und Atomare Physik, Technische Universit\"{a}t Berlin, Hardenbergstra{\ss}e 36, 10623 Berlin, Germany}

\author{L.~Leppert}
 \email{linn.leppert@uni-bayreuth.de}
 \affiliation{Theoretische Physik IV, Universit\"{a}t Bayreuth, 95440 Bayreuth, Germany}

\author{K.~Hirsch}
\author{A.~Langenberg}
\author{J.~Rittmann}
\author{M.~Kossick}
\author{M.~Vogel}
 \affiliation{Institut f\"{u}r Methoden und Instrumentierung der Synchrotronstrahlung, Helmholtz-Zentrum Berlin f\"{u}r Materialien und Energie GmbH, Albert-Einstein-Stra{\ss}e 15, 12489 Berlin, Germany}
 \affiliation{Institut f\"{u}r Optik und Atomare Physik, Technische Universit\"{a}t Berlin, Hardenbergstra{\ss}e 36, 10623 Berlin, Germany}

\author{R.~Richter}
 \affiliation{Institut f\"{u}r Optik und Atomare Physik, Technische Universit\"{a}t Berlin, Hardenbergstra{\ss}e 36, 10623 Berlin, Germany}

\author{A.~Terasaki}
 \affiliation{Cluster Research Laboratory, Toyota Technological Institute, 717-86 Futamata, Ichikawa, Chiba 272-0001, Japan}
 \affiliation{Department of Chemistry, Kyushu University, 6-10-1 Hakozaki, Higashi-ku, Fukuoka 812-8581, Japan}

\author{T.~M\"{o}ller}
 \affiliation{Institut f\"{u}r Optik und Atomare Physik, Technische Universit\"{a}t Berlin, Hardenbergstra{\ss}e 36, 10623 Berlin, Germany}

\author{B.~v.~Issendorff}
 \affiliation{Fakult\"{a}t f\"{u}r Physik, Universit\"{a}t Freiburg, Stefan-Meier-Stra{\ss}e 21, 79104 Freiburg, Germany}
 
 \author{S.~K\"{u}mmel}
 \affiliation{Theoretische Physik IV, Universit\"{a}t Bayreuth, 95440 Bayreuth, Germany}

\author{J.~T.~Lau}
 \email{tobias.lau@helmholtz-berlin.de}
 \affiliation{Institut f\"{u}r Methoden und Instrumentierung der Synchrotronstrahlung, Helmholtz-Zentrum Berlin f\"{u}r Materialien und Energie GmbH, Albert-Einstein-Stra{\ss}e 15, 12489 Berlin, Germany}

\date{\today}

\begin{abstract}
The interaction of a single manganese impurity with silicon is analyzed in a combined experimental and theoretical study of the electronic, magnetic, and structural properties of  manganese-doped silicon clusters. The structural transition from exohedral to endohedral doping coincides with a quenching of high-spin states. For all geometric structures investigated, we find a similar dependence of the magnetic moment on the manganese coordination number and nearest neighbor distance. This observation can be generalized to manganese point defects in bulk silicon, whose magnetic moments fall within the observed magnetic-to-nonmagnetic transition, and which therefore react very sensitively to changes in the local geometry. The results indicate that high spin states in manganese-doped silicon could be stabilized by an appropriate lattice expansion.
\end{abstract}


\maketitle
The interaction of a deliberately introduced impurity with a semiconductor material is one of the most fundamental problems of semiconductor physics. 
For magnetic impurities, an important question is the survival or quenching of the magnetic moment. From the pioneering studies of Ludwig and Woodbury \cite{Ludwig62} up to the present day, a wealth of experimental and theoretical studies have therefore been devoted to magnetic properties of transition metal doped semiconductors \cite{Beeler90,Wu07,Zhang08,Sato10,Zeng10}, semiconducting nanocrystals \cite{Huang05,Kuwen09,*Leitsmann09,Chen11}, and clusters \cite{Khanna02,Kumar03,Zheng05,Ngan12,Palagin12}.\@ 
A possible correlation between local magnetic moment and coordination number of the impurity has been noticed in theoretical work \cite{Zeng10}, but is challenging to investigate in bulk samples because of inhomogeneities, coalescence, or impurity band formation. These difficulties can be overcome by utilizing size-selected, singly-doped clusters as model systems where a transition metal atom occupies a well-defined position in the silicon host, without any interaction between impurities. 
Here, we study $\mathrm{MnSi}_n^+$ by x-ray absorption and x-ray magnetic circular dichroism (XMCD) spectroscopy of size-selected free clusters \cite{Lau08,Lau09a,Peredkov11,Niemeyer12,Vogel12} as a local and element-specific probe of electronic structure and magnetic properties. These experimental techniques are combined with non-empirical density functional theory (DFT) calculations. We find a clear dependence of the magnetic moment on the manganese coordination and nearest-neighbor distance. This result can be generalized to manganese point defects in bulk silicon.
\newline
Details of the experimental setup are given elsewhere \cite{Hirsch09,Niemeyer12}.\@ Very briefly, a continuous beam of $\mathrm{MnSi}_n^+$ clusters is produced in a magnetron gas aggregation source and transmitted through a combined radio-frequency hexapole ion guide and collision cell into a quadrupole mass filter. After mass selection, the clusters are accumulated in a cryogenic linear Paul trap and thermalized to $10 - 20$ K by collisions with helium buffer gas at $p \approx 10^{-3}~ \mathrm{mbar}$.\@ 
To study the local electronic and magnetic properties of $\mathrm{MnSi}_n^+$ by x-ray absorption and XMCD spectroscopy, a tunable monochromatic x-ray beam delivered by an undulator beamline at the synchrotron radiation facility BESSY II is coupled on-axis into the ion trap for resonant excitation at the manganese $L_{2,3}$-edge. This creates $\mathrm{Mn}^{+}$ and $\mathrm{Si}^{+}_2$ photoions, which are detected by a reflectron time-of-flight mass spectrometer. The incident photon energy is scanned from 618 - 686 eV to record photoion yield spectra that are a measure of the x-ray absorption cross section. 
For XMCD spectroscopy, which requires alignment of the total magnetic moment of free $\mathrm{MnSi}_n^+$, the liquid-helium cooled ion trap is placed inside the homogeneous magnetic field ($B = 5~\mathrm{T}$) of a superconducting solenoid, and ion yield spectra are recorded for parallel and antiparallel alignment of photon helicity and magnetic field \cite{Thole92,*Carra93}.\@ 
\begin{figure}
\includegraphics[width=\columnwidth]{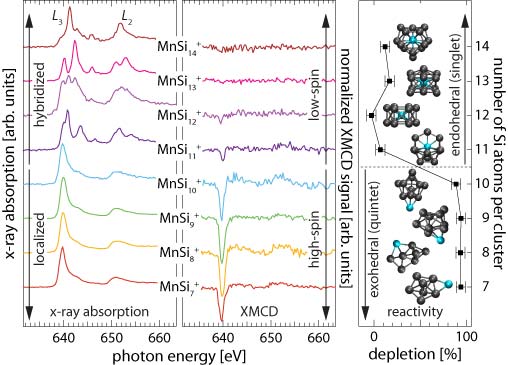}
\caption{\label{fig:XSpectra}(color online) Manganese $2p$ x-ray absorption (left) and XMCD (center) spectra of $\mathrm{MnSi}_n^+$ clusters ($n = 7 - 14$), indicating quenched magnetic moments for $n \ge 11$; corresponding ground state structures of $\mathrm{MnSi}_n^+$ and the depletion of singly doped clusters in the presence of $\mathrm{O}_2$ as a measure of the exohedral-to-endohedral transition (right).\@}
\end{figure}
\newline
In addition to magnetic and electronic properties, structural properties of $\mathrm{MnSi}_n^+$ are investigated. Similar to the reactivity and adsorption studies of doped silicon clusters by \citeauthor{Ohara03} \cite{Ohara03} and \citeauthor{Janssens07} \cite{Janssens07}, the exohedral-to-endohedral transition of manganese-doped silicon cluster cations is monitored via the depletion of $\mathrm{MnSi}_n^+$ in the cluster beam when introducing $p \approx 10^{-3}~ \mathrm{mbar}$ partial pressure of oxygen reactant gas into the hexapole collision cell. As can be seen in Fig.\ \ref{fig:XSpectra}, the depletion of singly doped $\mathrm{MnSi}_n^+$ is $89 - 94~\%$ for $n = 7 - 10$ but drops to $0 - 15~\%$ for $n \ge 11$.\@ This is due to the large difference of manganese and silicon reactivity towards oxygen that makes this depletion study a highly sensitive measure of the exohedral-to-endohedral transition, which takes place from $\mathrm{MnSi}_{10}^+$ to $\mathrm{MnSi}_{11}^+$.\@ 
\newline
This structural transition coincides with a marked change in the electronic properties of $\mathrm{MnSi}_n^+$ as can be seen in the manganese $L_{2,3}$ x-ray absorption and XMCD spectra. These probe local $2p \rightarrow 3d$ transitions at the manganese dopant and therefore reflect its electronic structure and magnetic moment. In Fig.\ \ref{fig:XSpectra}, exohedral clusters with $n = 7 - 10$ show nearly identical x-ray absorption spectra that indicate a very similar electronic structure of the manganese dopant.
In contrast, the x-ray absorption spectrum and thus the local electronic structure is more complex and varies strongly with the number of silicon atoms for endohedrally doped $\mathrm{MnSi}_n^+$.\@ Yet more striking, exohedral $\mathrm{MnSi}_n^+$ shows a pronounced XMCD asymmetry that vanishes for endohedral species. Even without applying XMCD sum rules \cite{Thole92,Carra93}, the
 XMCD asymmetry  is a qualitative and direct probe of magnetism and clearly indicates that manganese in exohedrally doped silicon clusters carries a magnetic moment which is quenched upon encapsulation. 
The residual XMCD asymmetry, which is observed for $n = 11$ and 12, is assigned to a slight contamination with $< 5~\%$ of $\mathrm{Mn}_{2}\mathrm{Si}_{n-2}^+$, because this XMCD signal and the corresponding lines in the x-ray absorption spectrum were tested to be proportional to the amount of $\mathrm{Mn}_{2}\mathrm{Si}_{n-2}^+$ that was observed simultaneously in mass spectrometry. 
\begin{figure}
\includegraphics[width=\columnwidth]{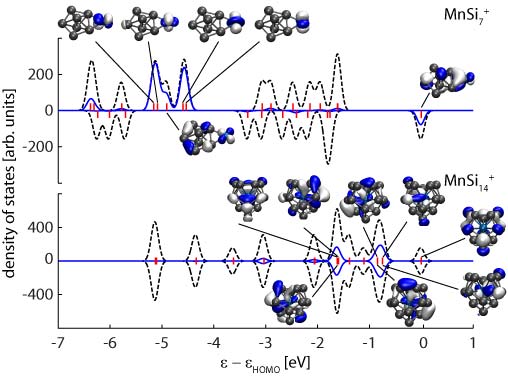}
\caption{\label{fig:DOS} (color online) Eigenvalue spectrum (bars), total (dotted line), and local Mn $3d$-projected (solid line) DOS with isosurface plots (at $\pm$ 0.04 a.u.) of the Mn $3d$ orbitals and the highest occupied orbital for $\mathrm{MnSi}_7^+$ and $\mathrm{MnSi}_{14}^+$. Positive (negative) values represent the spin-up (spin-down) channel. The $d$-like orbitals are localized on the manganese atom in $\mathrm{MnSi}_7^+$, whereas they are delocalized in $\mathrm{MnSi}_{14}^+$.\@}
\end{figure}
\newline
To further analyze the exohedral-to-endohedral and magnetic-to-nonmagnetic transition in $\mathrm{MnSi}_n^+$, we performed a thorough global geometry optimization in a simulated annealing \cite{Kirkpatrick1983} and modified "big bang" \cite{Jackson04} approach, details of which are given in the Supplemental Material \cite{SM_MnSi}.\@ 
The calculations were carried out in a DFT framework using the Perdew-Burke-Ernzerhof one-parameter hybrid (PBE0) \cite{Adamo99} as implemented in {\sc{turbomole}} \cite{TURBOMOLE3}.\@ The PBE0 exchange correlation (xc) functional was chosen because it partly cancels the effects of the self-interaction error \cite{Perdew81} that is inherent in commonly used semilocal functionals and often leads to erroneous results for the electronic structure of systems in which the highest occupied orbitals differ significantly in their degree of spatial localization \cite{Korzdorfer09,*Korzdorfer10b}.\@ This is particularly true for systems containing transition metal elements, such as $\mathrm{MnSi}_n^+$.\@ 
\newline
The assigned ground state structures of $\mathrm{MnSi}_n^+$ that result from our calculations are depicted in Fig.\ \ref{fig:XSpectra}.\@ 
With the exception of $\mathrm{MnSi}_8^+$, the predicted structures of exohedrally doped $\mathrm{MnSi}_n^+$ clusters for $n = 7 - 10$ correspond to those of $\mathrm{Si}_{n+1}^+$ (cf.\ Refs.\ \onlinecite{Lyon09, Vogel12}), where one silicon atom is replaced by manganese. 
In contrast, no structural similarity to the corresponding $\mathrm{Si}_{n+1}^+$ clusters can be observed in the endohedral size regime, where manganese is encapsulated by silicon. 
These structural findings agree qualitatively with results of the DFT and infrared spectroscopy study by \citeauthor{Ngan12} \cite{Ngan12}.\@ 
Moreover, our calculations show that the magnetic moment of $\mathrm{MnSi}_n^+$ is quenched from $4~\mu_B$ to $0~\mu_B$ at the exohedral-to-endohedral transition, in perfect agreement with the experimentally observed disappearance of the XMCD signal  from $n \le 10$  to $n \ge 11$.\@ 
\begin{figure}
\includegraphics[width=\columnwidth]{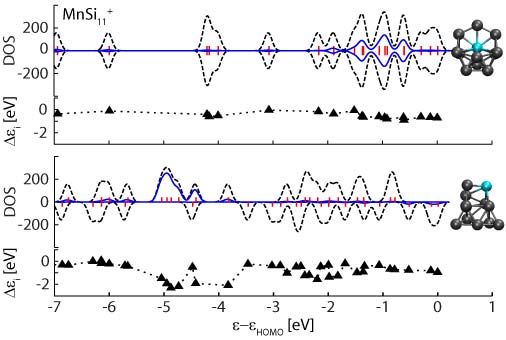}
\caption{\label{fig:sie} (color online) Eigenvalue spectrum (bars), total DOS (dotted line), and manganese-projected local DOS (solid line) of the true endohedral ground state of $\mathrm{MnSi}_{11}^+$ (top) and of the false exohedral ground state predicted by PBE0 (bottom).
$\Delta \varepsilon_i$ is a measure for how much the PBE0 eigenvalues are affected by the self-interaction error (see text).}
\end{figure}
\newline
The change in local electronic structure of the manganese dopant that was observed in the x-ray absorption spectra at the structural transition of $\mathrm{MnSi}_n^+$ in Fig.\ref{fig:XSpectra} is reflected in the occupied eigenvalue spectrum, which is shown in its usual interpretation as a density of states (DOS) in Fig.\ \ref{fig:DOS} for $\mathrm{MnSi}_{7}^+$ and $\mathrm{MnSi}_{14}^+$, representing the exohedral and the endohedral size regime. 
For $\mathrm{MnSi}_7^+$, the occupied states of manganese $3d$ character are mostly isolated at $\approx 4 - 5$ eV below the Fermi level and are tightly localized at the manganese dopant, i.e., they largely preserve a (perturbed) atomic character and only weakly interact with silicon states, as can be seen from the DOS and the isosurface plots. 
Because of this weak interaction, the localized orbitals of manganese $3d$ character are qualitatively very similar for all exohedral $\mathrm{MnSi}_n^+$ clusters and lead to the nearly identical x-ray absorption spectra in Fig.\  \ref{fig:XSpectra}.
This notion of, at least partly, atomic manganese $3d$ states is lost in the endohedral size regime, represented by $\mathrm{MnSi}_{14}^+$ in Fig.\ \ref{fig:DOS}.\@ Here, orbitals with partial manganese $3d$ character are strongly hybridized with silicon states and are shifted to $\approx 0.5 - 2$ eV below the Fermi level, i.e., they participate strongly in bonding and are delocalized over the silicon frame. Consequently, the manganese $3d$ derived partial DOS sensitively depends on the structure of the silicon cage, which is reflected in the variation of the x-ray absorption spectra of endohedral $\mathrm{MnSi}_n^+$ for different $n$ in Fig.\  \ref{fig:XSpectra}.\@ 
As a result of strong $spd$ hybridization with the participation of all manganese valence orbitals, the magnetic moment is completely quenched in endohedrally doped clusters.
\begin{figure}
\includegraphics[width=\columnwidth]{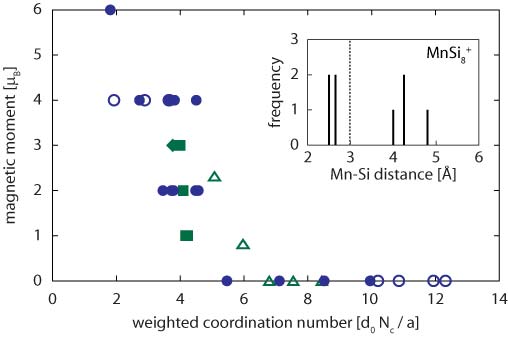}
\caption{\label{fig:coordMoment} (color online) Magnetic moment versus weighted coordination $d_0 N_c/a$ for ground state (open circles) and higher energy (solid circles) isomers of $\mathrm{MnSi}_n^+$ ($n = 7 - 14$), isolated neutral Mn impurities in bulk (\cite{Zhang08}, solid squares) and amorphous silicon (\cite{Zeng10}, open triangles), and in a silicon nanocrystal (\cite{Leitsmann09}, solid diamond).\@ Inset: bond length distribution of $\mathrm{MnSi}_8^+$ with first coordination sphere (dotted line).\@}
\end{figure}
\newline
Because of the well known self-interaction error \cite{Perdew81}, a treatment of this change from a partly localized, atomic-like situation to one in which all orbitals are delocalized on similar length scales poses serious difficulties to standard approximations of the xc functional. Therefore, the DFT results have to be analyzed carefully at the structural transition around $\mathrm{MnSi}_{11}^+$, where exohedral and endohedral isomers can be expected to be closest in energy. 
For $\mathrm{MnSi}_{11}^+$, PBE0 predicts an exohedral ground state with a total magnetic moment of $2~\mu_B$, which is 1.13 eV lower in energy than an endohedral isomer that would be in agreement with the experimental results. We attribute this discrepancy to uncertainties in theory for two reasons: 
First, we argue that the experimentally observed endohedral structure is indeed the ground state since we have observed ground-state structures of the related systems $\mathrm{Si}_n^+$ and $\mathrm{VSi}_n^+$ \cite{Lau09a,Lau11,Vogel12} under similar experimental conditions and for comparable size ranges. Second, a closer look at the theoretical result reveals that the failure of PBE0 for $\mathrm{MnSi}_{11}^+$ can indeed be explained in terms of the effect of self-interaction on the eigenvalue spectrum of both isomers. The self-interaction error $e_i$ of the orbital $\varphi_i$,
\begin{equation*}
 e_i = \left\langle \varphi_i \left| v_H \left[ | \varphi_i |^2 \right] + v_{xc}^{\text{approx}}\left[ | \varphi_i| ^2,0 \right] \right| \varphi_i \right\rangle,
\end{equation*}
can be used to quantify the reliability of the eigenvalue spectrum \cite{Korzdorfer09,*Korzdorfer10b}.\@ Here $v_H$ is the electrostatic Hartree-potential and $v_{xc}^{\text{approx}}$ is an approximate xc potential. Details of the calculation of $e_i$ are given in the Supplemental Material \cite{SM_MnSi} and references therein \cite{Kronik2006,Perdew1996,Korzdorfer10a}.\@ The self-interaction corrected eigenvalues $\varepsilon_i$ can be estimated as $\varepsilon_i \approx \varepsilon_i^{\text{approx}} - e_i$ \cite{Perdew81}, where $\varepsilon_i^{\text{approx}}$ results from a self-consistent calculation using $v_{xc}^{\text{approx}}$.\@ Fig.\ \ref{fig:sie} compares the difference $\Delta \varepsilon_i$ between $\varepsilon_i$ and the PBE0 eigenvalues $\varepsilon_i^{PBE0}$ as a measure of how reliably PBE0 cancels the orbital self-interaction error in the two $\mathrm{MnSi}_{11}^+$ isomers. 
The large value of $\Delta \varepsilon_i$ as well as its scatter shows that the exohedral PBE0 result is nonreliable in the case of $\mathrm{MnSi}_{11}^+$ because self-interaction strongly affects orbitals that participate in bonding. For the endohedral isomer with delocalized orbitals, the PBE0 eigenvalues are almost identical to $\varepsilon_i$, and are thus reliable. In view of these arguments, we conclude that $\mathrm{MnSi}_{11}^+$ is the smallest endohedral structure.
This reassessment of the energetic order does not alter the observed quenching of the magnetic moment at the exohedral-to-endohedral transition since every endohedral isomer of $\mathrm{MnSi}_{n}^+$ that resulted from our calculations is predicted to adopt a singlet state by PBE0.\@ We therefore stress that these deliberations only lead us to reconsider the predicted critical size in accordance with our reactivity studies, but do not change the electronic or magnetic properties.
\newline
The structural change at the exohedral-to-endohedral transition of $\mathrm{MnSi}_n^+$ can be quantified by the coordination number $N_c$ of manganese, i.e., the number of silicon atoms in the first coordination sphere as exemplified for $\mathrm{MnSi}_8^+$ in the inset of Fig.\ \ref{fig:coordMoment}.\@ In exohedral clusters, manganese adopts a minimal coordination number of $N_c = 2 \mathrm{-} 4$, while in endohedral clusters $N_c$ is maximized to $11 \mathrm{-} 14$, i.e., all silicon atoms are within the first coordination sphere of manganese. 
The average Mn-Si bond length $a$ elucidates why encapsulation of the manganese dopant becomes energetically favorable only for $n \ge 11$: In the ground state structures, the Mn-Si nearest neighbor distance expands from $a = 2.43 - 2.58$ \AA{} in exohedral to $a = 2.53 - 2.67$ \AA{} in endohedral clusters. In contrast, it would be compressed to $a = 2.35$ \AA{} in the higher energy endohedral isomer of $\mathrm{MnSi}_{10}^+$.\@ Even though manganese favors high coordination in silicon \cite{Wu04a}, this strain, which becomes even more pronounced in smaller clusters, precludes endohedral ground states for $n \le 10$.\@
\newline
The abrupt change in coordination at the structural transition is interrelated with the quenching of the magnetic moment as illustrated in Fig.\ \ref{fig:coordMoment}: Here, the calculated magnetic moments of  $\mathrm{MnSi}_n^+$ are plotted versus the weighted coordination number $d_0 N_c / a$, which takes into account both the number of silicon nearest neighbors $N_c$ as well as their average distance $a$ to the manganese atom, normalized to the nearest neighbor distance $d_0$ in bulk silicon.\@  
Low-coordinated exohedral clusters with $d_0 N_c / a = 1.9 - 3.7$ $(N_c = 2 - 4)$ carry a magnetic moment of $4~\mu_\mathrm{B}$, which is quenched to $0~\mu_\mathrm{B}$ in high-coordinated species with $d_0 N_c / a = 10.2 - 12.3$ $(N_c = 11 - 14)$.\@
This relation of magnetic moment and weighted coordination also holds for higher-energy isomers that are included in Fig.\ \ref{fig:coordMoment} and mark the transition from magnetic to nonmagnetic impurities around $d_0 N_c / a \approx 4$.\@ 
\newline
Fig.\ \ref{fig:coordMoment} shows that this observation can be generalized to extended systems, i.e., to a neutral manganese impurity at a substitutional site in crystalline silicon \cite{Zhang08} or in hydrogen-passivated silicon nanocrystals \cite{Leitsmann09}.\@ It also applies to very low concentrations of manganese in amorphous silicon, for which a possible relation between magnetic moment and coordination has been pointed out \cite{Zeng10}.\@ However, although $N_c$ is the leading term, it does not account for the dependence of the local magnetic moment on the nearest-neighbor distance \cite{Zhang08} that becomes important around $N_c = 4$ and is included in  $d_0 N_c / a$.\@ As can be seen in Fig.\ \ref{fig:coordMoment}, manganese-doped bulk silicon is just at the transition from high-spin to low-spin states and therefore reacts very sensitively to changes in $d_0 N_c / a$.\@ This might explain the large scatter in experimental results on manganese-doped silicon and indicates that high-spin states could be stabilized by an appropriate expansion of the lattice parameter, e.g., in ultrathin films or passivated nanocrystals.  
\newline
In summary, the magnetic moment of manganese-doped silicon has been investigated over a wide range of structural parameters, including extreme coordination numbers from 2 - 14.\@ The study of singly doped, size-selected $\mathrm{MnSi}_n^+$ clusters avoids impurity-band formation or interaction between impurities that might be present in experiments on bulk samples, but also in calculations with periodic boundary conditions. We are thus able to show that the observed quenching of the magnetic moment is not a result of impurity band formation but of the electronic interaction with the silicon host. A universal correlation of the magnetic moment and the weighted coordination number is observed, providing guidelines to the stabilization of high-spin states in dilute manganese-doped silicon.
\newline
This work was supported by DFG grant No.\ LA 2398/5-1 within FOR 1282.\@ Beamtime for this project was granted at BESSY II beamlines UE52-SGM and U49/2-PGM-1, operated by Helmholtz-Zentrum Berlin. The superconducting magnet was provided by Toyota Technological Institute. SK and LL acknowledge financial support by DFG SFB 840. SK additionally acknowledges support by the GIF.\@ We thank V.\ Forster for providing the code for random coordinate generation, and E.\ Janssens for the experimental IRMPD spectra. AT acknowledges financial support
by Genesis Research Institute, Inc. BvI acknowledges travel support by HZB.\@


\begin{thebibliography}{39}%
\makeatletter
\providecommand \@ifxundefined [1]{%
 \@ifx{#1\undefined}
}%
\providecommand \@ifnum [1]{%
 \ifnum #1\expandafter \@firstoftwo
 \else \expandafter \@secondoftwo
 \fi
}%
\providecommand \@ifx [1]{%
 \ifx #1\expandafter \@firstoftwo
 \else \expandafter \@secondoftwo
 \fi
}%
\providecommand \natexlab [1]{#1}%
\providecommand \enquote  [1]{``#1''}%
\providecommand \bibnamefont  [1]{#1}%
\providecommand \bibfnamefont [1]{#1}%
\providecommand \citenamefont [1]{#1}%
\providecommand \href@noop [0]{\@secondoftwo}%
\providecommand \href [0]{\begingroup \@sanitize@url \@href}%
\providecommand \@href[1]{\@@startlink{#1}\@@href}%
\providecommand \@@href[1]{\endgroup#1\@@endlink}%
\providecommand \@sanitize@url [0]{\catcode `\\12\catcode `\$12\catcode
  `\&12\catcode `\#12\catcode `\^12\catcode `\_12\catcode `\%12\relax}%
\providecommand \@@startlink[1]{}%
\providecommand \@@endlink[0]{}%
\providecommand \url  [0]{\begingroup\@sanitize@url \@url }%
\providecommand \@url [1]{\endgroup\@href {#1}{\urlprefix }}%
\providecommand \urlprefix  [0]{URL }%
\providecommand \Eprint [0]{\href }%
\@ifxundefined \urlstyle {%
  \providecommand \doi  [0]{\begingroup \@sanitize@url \@doi}%
  \providecommand \@doi [1]{\endgroup \@@startlink {\doibase
  #1}doi:\discretionary {}{}{}#1\@@endlink }%
}{%
  \providecommand \doi  [0]{doi:\discretionary{}{}{}\begingroup
  \urlstyle{rm}\Url }%
}%
\providecommand \doibase [0]{http://dx.doi.org/}%
\providecommand \Doi [0]{\begingroup \@sanitize@url \@Doi }%
\providecommand \@Doi  [1]{\endgroup\@@startlink{\doibase#1}\@@Doi}%
\providecommand \@@Doi [1]{#1\@@endlink}%
\providecommand \selectlanguage [0]{\@gobble}%
\providecommand \bibinfo  [0]{\@secondoftwo}%
\providecommand \bibfield  [0]{\@secondoftwo}%
\providecommand \translation [1]{[#1]}%
\providecommand \BibitemOpen [0]{}%
\providecommand \bibitemStop [0]{}%
\providecommand \bibitemNoStop [0]{.\EOS\space}%
\providecommand \EOS [0]{\spacefactor3000\relax}%
\providecommand \BibitemShut  [1]{\csname bibitem#1\endcsname}%
%</preamble>
\bibitem [{\citenamefont {Ludwig}\ and\ \citenamefont
  {Woodbury}(1962)}]{Ludwig62}%
  \BibitemOpen
  \bibfield  {author} {\bibinfo {author} {\bibfnamefont {G.~W.}\ \bibnamefont
  {Ludwig}}\ and\ \bibinfo {author} {\bibfnamefont {H.~H.}\ \bibnamefont
  {Woodbury}},\ }\Doi {10.1016/S0081-1947(08)60458-0} {\bibfield  {journal}
  {\bibinfo  {journal} {Solid State Phys.},\ }\textbf {\bibinfo {volume}
  {13}},\ \bibinfo {pages} {223 } (\bibinfo {year} {1962})}\BibitemShut
  {NoStop}%
\bibitem [{\citenamefont {Beeler}\ \emph {et~al.}(1990)\citenamefont {Beeler},
  \citenamefont {Andersen},\ and\ \citenamefont {Scheffler}}]{Beeler90}%
  \BibitemOpen
  \bibfield  {author} {\bibinfo {author} {\bibfnamefont {F.}~\bibnamefont
  {Beeler}}, \bibinfo {author} {\bibfnamefont {O.~K.}\ \bibnamefont
  {Andersen}}, \ and\ \bibinfo {author} {\bibfnamefont {M.}~\bibnamefont
  {Scheffler}},\ }\href@noop {} {\bibfield  {journal} {\bibinfo  {journal}
  {Phys. Rev. B},\ }\textbf {\bibinfo {volume} {41}},\ \bibinfo {pages} {1603}
  (\bibinfo {year} {1990})}\BibitemShut {NoStop}%
\bibitem [{\citenamefont {Wu}\ \emph {et~al.}(2007)\citenamefont {Wu},
  \citenamefont {Kratzer},\ and\ \citenamefont {Scheffler}}]{Wu07}%
  \BibitemOpen
  \bibfield  {author} {\bibinfo {author} {\bibfnamefont {H.}~\bibnamefont
  {Wu}}, \bibinfo {author} {\bibfnamefont {P.}~\bibnamefont {Kratzer}}, \ and\
  \bibinfo {author} {\bibfnamefont {M.}~\bibnamefont {Scheffler}},\ }\href@noop
  {} {\bibfield  {journal} {\bibinfo  {journal} {Phys. Rev. Lett.},\ }\textbf
  {\bibinfo {volume} {98}},\ \bibinfo {pages} {117202} (\bibinfo {year}
  {2007})}\BibitemShut {NoStop}%
\bibitem [{\citenamefont {Zhang}\ \emph {et~al.}(2008)\citenamefont {Zhang},
  \citenamefont {Partoens}, \citenamefont {Chang},\ and\ \citenamefont
  {Peeters}}]{Zhang08}%
  \BibitemOpen
  \bibfield  {author} {\bibinfo {author} {\bibfnamefont {Z.~Z.}\ \bibnamefont
  {Zhang}}, \bibinfo {author} {\bibfnamefont {B.}~\bibnamefont {Partoens}},
  \bibinfo {author} {\bibfnamefont {K.}~\bibnamefont {Chang}}, \ and\ \bibinfo
  {author} {\bibfnamefont {F.~M.}\ \bibnamefont {Peeters}},\ }\href@noop {}
  {\bibfield  {journal} {\bibinfo  {journal} {Phys. Rev. B},\ }\textbf
  {\bibinfo {volume} {77}},\ \bibinfo {pages} {155201} (\bibinfo {year}
  {2008})}\BibitemShut {NoStop}%
\bibitem [{\citenamefont {Sato}\ \emph {et~al.}(2010)\citenamefont {Sato},
  \citenamefont {Bergqvist}, \citenamefont {Kudrnovsky}, \citenamefont
  {Dederichs}, \citenamefont {Eriksson}, \citenamefont {Turek}, \citenamefont
  {Sanyal}, \citenamefont {Bouzerar}, \citenamefont {Katayama-Yoshida},
  \citenamefont {Dinh}, \citenamefont {Fukushima}, \citenamefont {Kizaki},\
  and\ \citenamefont {Zeller}}]{Sato10}%
  \BibitemOpen
  \bibfield  {author} {\bibinfo {author} {\bibfnamefont {K.}~\bibnamefont
  {Sato}}, \bibinfo {author} {\bibfnamefont {L.}~\bibnamefont {Bergqvist}},
  \bibinfo {author} {\bibfnamefont {J.}~\bibnamefont {Kudrnovsky}}, \bibinfo
  {author} {\bibfnamefont {P.~H.}\ \bibnamefont {Dederichs}}, \bibinfo {author}
  {\bibfnamefont {O.}~\bibnamefont {Eriksson}}, \bibinfo {author}
  {\bibfnamefont {I.}~\bibnamefont {Turek}}, \bibinfo {author} {\bibfnamefont
  {B.}~\bibnamefont {Sanyal}}, \bibinfo {author} {\bibfnamefont
  {G.}~\bibnamefont {Bouzerar}}, \bibinfo {author} {\bibfnamefont
  {H.}~\bibnamefont {Katayama-Yoshida}}, \bibinfo {author} {\bibfnamefont
  {V.~A.}\ \bibnamefont {Dinh}}, \bibinfo {author} {\bibfnamefont
  {T.}~\bibnamefont {Fukushima}}, \bibinfo {author} {\bibfnamefont
  {H.}~\bibnamefont {Kizaki}}, \ and\ \bibinfo {author} {\bibfnamefont
  {R.}~\bibnamefont {Zeller}},\ }\href@noop {} {\bibfield  {journal} {\bibinfo
  {journal} {Rev. Mod. Phys.},\ }\textbf {\bibinfo {volume} {82}},\ \bibinfo
  {pages} {1633} (\bibinfo {year} {2010})}\BibitemShut {NoStop}%
\bibitem [{\citenamefont {Zeng}\ \emph {et~al.}(2010)\citenamefont {Zeng},
  \citenamefont {Cao}, \citenamefont {Helgren}, \citenamefont {Karel},
  \citenamefont {Arenholz}, \citenamefont {Ouyang}, \citenamefont {Smith},
  \citenamefont {Wu},\ and\ \citenamefont {Hellman}}]{Zeng10}%
  \BibitemOpen
  \bibfield  {author} {\bibinfo {author} {\bibfnamefont {L.}~\bibnamefont
  {Zeng}}, \bibinfo {author} {\bibfnamefont {J.~X.}\ \bibnamefont {Cao}},
  \bibinfo {author} {\bibfnamefont {E.}~\bibnamefont {Helgren}}, \bibinfo
  {author} {\bibfnamefont {J.}~\bibnamefont {Karel}}, \bibinfo {author}
  {\bibfnamefont {E.}~\bibnamefont {Arenholz}}, \bibinfo {author}
  {\bibfnamefont {L.}~\bibnamefont {Ouyang}}, \bibinfo {author} {\bibfnamefont
  {D.~J.}\ \bibnamefont {Smith}}, \bibinfo {author} {\bibfnamefont {R.~Q.}\
  \bibnamefont {Wu}}, \ and\ \bibinfo {author} {\bibfnamefont {F.}~\bibnamefont
  {Hellman}},\ }\href@noop {} {\bibfield  {journal} {\bibinfo  {journal} {Phys.
  Rev. B},\ }\textbf {\bibinfo {volume} {82}},\ \bibinfo {pages} {165202}
  (\bibinfo {year} {2010})}\BibitemShut {NoStop}%
\bibitem [{\citenamefont {Huang}\ \emph {et~al.}(2005)\citenamefont {Huang},
  \citenamefont {Makmal}, \citenamefont {Chelikowsky},\ and\ \citenamefont
  {Kronik}}]{Huang05}%
  \BibitemOpen
  \bibfield  {author} {\bibinfo {author} {\bibfnamefont {X.}~\bibnamefont
  {Huang}}, \bibinfo {author} {\bibfnamefont {A.}~\bibnamefont {Makmal}},
  \bibinfo {author} {\bibfnamefont {J.~R.}\ \bibnamefont {Chelikowsky}}, \ and\
  \bibinfo {author} {\bibfnamefont {L.}~\bibnamefont {Kronik}},\ }\href@noop {}
  {\bibfield  {journal} {\bibinfo  {journal} {Phys. Rev. Lett.},\ }\textbf
  {\bibinfo {volume} {94}},\ \bibinfo {pages} {236801} (\bibinfo {year}
  {2005})}\BibitemShut {NoStop}%
\bibitem [{\citenamefont {K{\"u}wen}\ \emph {et~al.}(2009)\citenamefont
  {K{\"u}wen}, \citenamefont {Leitsmann},\ and\ \citenamefont
  {Bechstedt}}]{Kuwen09}%
  \BibitemOpen
  \bibfield  {author} {\bibinfo {author} {\bibfnamefont {F.}~\bibnamefont
  {K{\"u}wen}}, \bibinfo {author} {\bibfnamefont {R.}~\bibnamefont
  {Leitsmann}}, \ and\ \bibinfo {author} {\bibfnamefont {F.}~\bibnamefont
  {Bechstedt}},\ }\href@noop {} {\bibfield  {journal} {\bibinfo  {journal}
  {Phys. Rev. B},\ }\textbf {\bibinfo {volume} {80}},\ \bibinfo {pages}
  {045203} (\bibinfo {year} {2009})}\BibitemShut {NoStop}%
\bibitem [{\citenamefont {Leitsmann}\ \emph {et~al.}(2009)\citenamefont
  {Leitsmann}, \citenamefont {Panse}, \citenamefont {K{\"u}wen},\ and\
  \citenamefont {Bechstedt}}]{Leitsmann09}%
  \BibitemOpen
  \bibfield  {author} {\bibinfo {author} {\bibfnamefont {R.}~\bibnamefont
  {Leitsmann}}, \bibinfo {author} {\bibfnamefont {C.}~\bibnamefont {Panse}},
  \bibinfo {author} {\bibfnamefont {F.}~\bibnamefont {K{\"u}wen}}, \ and\
  \bibinfo {author} {\bibfnamefont {F.}~\bibnamefont {Bechstedt}},\ }\href@noop
  {} {\bibfield  {journal} {\bibinfo  {journal} {Phys. Rev. B},\ }\textbf
  {\bibinfo {volume} {80}},\ \bibinfo {pages} {104412} (\bibinfo {year}
  {2009})}\BibitemShut {NoStop}%
\bibitem [{\citenamefont {Chen}\ \emph {et~al.}(2011)\citenamefont {Chen},
  \citenamefont {Pi},\ and\ \citenamefont {Yang}}]{Chen11}%
  \BibitemOpen
  \bibfield  {author} {\bibinfo {author} {\bibfnamefont {X.}~\bibnamefont
  {Chen}}, \bibinfo {author} {\bibfnamefont {X.}~\bibnamefont {Pi}}, \ and\
  \bibinfo {author} {\bibfnamefont {D.}~\bibnamefont {Yang}},\ }\href@noop {}
  {\bibfield  {journal} {\bibinfo  {journal} {Appl. Phys. Lett.},\ }\textbf
  {\bibinfo {volume} {99}},\ \bibinfo {pages} {193108} (\bibinfo {year}
  {2011})}\BibitemShut {NoStop}%
\bibitem [{\citenamefont {Khanna}\ \emph {et~al.}(2002)\citenamefont {Khanna},
  \citenamefont {Rao},\ and\ \citenamefont {Jena}}]{Khanna02}%
  \BibitemOpen
  \bibfield  {author} {\bibinfo {author} {\bibfnamefont {S.~N.}\ \bibnamefont
  {Khanna}}, \bibinfo {author} {\bibfnamefont {B.~K.}\ \bibnamefont {Rao}}, \
  and\ \bibinfo {author} {\bibfnamefont {P.}~\bibnamefont {Jena}},\ }\href@noop
  {} {\bibfield  {journal} {\bibinfo  {journal} {Phys. Rev. Lett.},\ }\textbf
  {\bibinfo {volume} {89}},\ \bibinfo {pages} {016803} (\bibinfo {year}
  {2002})}\BibitemShut {NoStop}%
\bibitem [{\citenamefont {Kumar}\ and\ \citenamefont
  {Kawazoe}(2003)}]{Kumar03}%
  \BibitemOpen
  \bibfield  {author} {\bibinfo {author} {\bibfnamefont {V.}~\bibnamefont
  {Kumar}}\ and\ \bibinfo {author} {\bibfnamefont {Y.}~\bibnamefont
  {Kawazoe}},\ }\href@noop {} {\bibfield  {journal} {\bibinfo  {journal} {Appl.
  Phys. Lett.},\ }\textbf {\bibinfo {volume} {83}},\ \bibinfo {pages} {2677}
  (\bibinfo {year} {2003})}\BibitemShut {NoStop}%
\bibitem [{\citenamefont {Zheng}\ \emph {et~al.}(2005)\citenamefont {Zheng},
  \citenamefont {Nilles}, \citenamefont {Radisic},\ and\ \citenamefont {{Bowen,
  Jr.}}}]{Zheng05}%
  \BibitemOpen
  \bibfield  {author} {\bibinfo {author} {\bibfnamefont {W.}~\bibnamefont
  {Zheng}}, \bibinfo {author} {\bibfnamefont {J.~M.}\ \bibnamefont {Nilles}},
  \bibinfo {author} {\bibfnamefont {D.}~\bibnamefont {Radisic}}, \ and\
  \bibinfo {author} {\bibfnamefont {K.~H.}\ \bibnamefont {{Bowen, Jr.}}},\
  }\href@noop {} {\bibfield  {journal} {\bibinfo  {journal} {J. Chem. Phys.},\
  }\textbf {\bibinfo {volume} {122}},\ \bibinfo {pages} {071101} (\bibinfo
  {year} {2005})}\BibitemShut {NoStop}%
\bibitem [{\citenamefont {Ngan}\ \emph {et~al.}(2012)\citenamefont {Ngan},
  \citenamefont {Janssens}, \citenamefont {Claes}, \citenamefont {Lyon},
  \citenamefont {Fielicke}, \citenamefont {Nguyen},\ and\ \citenamefont
  {Lievens}}]{Ngan12}%
  \BibitemOpen
  \bibfield  {author} {\bibinfo {author} {\bibfnamefont {V.~T.}\ \bibnamefont
  {Ngan}}, \bibinfo {author} {\bibfnamefont {E.}~\bibnamefont {Janssens}},
  \bibinfo {author} {\bibfnamefont {P.}~\bibnamefont {Claes}}, \bibinfo
  {author} {\bibfnamefont {J.~T.}\ \bibnamefont {Lyon}}, \bibinfo {author}
  {\bibfnamefont {A.}~\bibnamefont {Fielicke}}, \bibinfo {author}
  {\bibfnamefont {M.~T.}\ \bibnamefont {Nguyen}}, \ and\ \bibinfo {author}
  {\bibfnamefont {P.}~\bibnamefont {Lievens}},\ }\Doi {10.1002/chem.201201839}
  {\bibfield  {journal} {\bibinfo  {journal} {Chem. Eur. J.},\ }\textbf
  {\bibinfo {volume} {18}},\ \bibinfo {pages} {15788} (\bibinfo {year}
  {2012})}\BibitemShut {NoStop}%
\bibitem [{\citenamefont {Palagin}\ and\ \citenamefont
  {Reuter}(2012)}]{Palagin12}%
  \BibitemOpen
  \bibfield  {author} {\bibinfo {author} {\bibfnamefont {D.}~\bibnamefont
  {Palagin}}\ and\ \bibinfo {author} {\bibfnamefont {K.}~\bibnamefont
  {Reuter}},\ }\href@noop {} {\bibfield  {journal} {\bibinfo  {journal} {Phys.
  Rev. B},\ }\textbf {\bibinfo {volume} {86}},\ \bibinfo {pages} {045416}
  (\bibinfo {year} {2012})}\BibitemShut {NoStop}%
\bibitem [{\citenamefont {Lau}\ \emph {et~al.}(2008)\citenamefont {Lau},
  \citenamefont {Rittmann}, \citenamefont {Zamudio-Bayer}, \citenamefont
  {Vogel}, \citenamefont {Hirsch}, \citenamefont {Klar}, \citenamefont
  {Lofink}, \citenamefont {M\"{o}ller},\ and\ \citenamefont
  {v.~Issendorff}}]{Lau08}%
  \BibitemOpen
  \bibfield  {author} {\bibinfo {author} {\bibfnamefont {J.~T.}\ \bibnamefont
  {Lau}}, \bibinfo {author} {\bibfnamefont {J.}~\bibnamefont {Rittmann}},
  \bibinfo {author} {\bibfnamefont {V.}~\bibnamefont {Zamudio-Bayer}}, \bibinfo
  {author} {\bibfnamefont {M.}~\bibnamefont {Vogel}}, \bibinfo {author}
  {\bibfnamefont {K.}~\bibnamefont {Hirsch}}, \bibinfo {author} {\bibfnamefont
  {P.}~\bibnamefont {Klar}}, \bibinfo {author} {\bibfnamefont {F.}~\bibnamefont
  {Lofink}}, \bibinfo {author} {\bibfnamefont {T.}~\bibnamefont {M\"{o}ller}},
  \ and\ \bibinfo {author} {\bibfnamefont {B.}~\bibnamefont {v.~Issendorff}},\
  }\Doi {10.1103/PhysRevLett.101.153401} {\bibfield  {journal} {\bibinfo
  {journal} {Phys. Rev. Lett.},\ }\textbf {\bibinfo {volume} {101}},\ \bibinfo
  {eid} {153401} (\bibinfo {year} {2008})}\BibitemShut {NoStop}%
\bibitem [{\citenamefont {Lau}\ \emph {et~al.}(2009)\citenamefont {Lau},
  \citenamefont {Hirsch}, \citenamefont {Klar}, \citenamefont {Langenberg},
  \citenamefont {Lofink}, \citenamefont {Richter}, \citenamefont {Rittmann},
  \citenamefont {Vogel}, \citenamefont {Zamudio-Bayer}, \citenamefont
  {M\"oller},\ and\ \citenamefont {v.~Issendorff}}]{Lau09a}%
  \BibitemOpen
  \bibfield  {author} {\bibinfo {author} {\bibfnamefont {J.~T.}\ \bibnamefont
  {Lau}}, \bibinfo {author} {\bibfnamefont {K.}~\bibnamefont {Hirsch}},
  \bibinfo {author} {\bibfnamefont {P.}~\bibnamefont {Klar}}, \bibinfo {author}
  {\bibfnamefont {A.}~\bibnamefont {Langenberg}}, \bibinfo {author}
  {\bibfnamefont {F.}~\bibnamefont {Lofink}}, \bibinfo {author} {\bibfnamefont
  {R.}~\bibnamefont {Richter}}, \bibinfo {author} {\bibfnamefont
  {J.}~\bibnamefont {Rittmann}}, \bibinfo {author} {\bibfnamefont
  {M.}~\bibnamefont {Vogel}}, \bibinfo {author} {\bibfnamefont
  {V.}~\bibnamefont {Zamudio-Bayer}}, \bibinfo {author} {\bibfnamefont
  {T.}~\bibnamefont {M\"oller}}, \ and\ \bibinfo {author} {\bibfnamefont
  {B.}~\bibnamefont {v.~Issendorff}},\ }\Doi {10.1103/PhysRevA.79.053201}
  {\bibfield  {journal} {\bibinfo  {journal} {Phys. Rev. A},\ }\textbf
  {\bibinfo {volume} {79}},\ \bibinfo {eid} {053201} (\bibinfo {year}
  {2009})}\BibitemShut {NoStop}%
\bibitem [{\citenamefont {Peredkov}\ \emph {et~al.}(2011)\citenamefont
  {Peredkov}, \citenamefont {Neeb}, \citenamefont {Eberhardt}, \citenamefont
  {Meyer}, \citenamefont {Tombers}, \citenamefont {Kampschulte},\ and\
  \citenamefont {Niedner-Schatteburg}}]{Peredkov11}%
  \BibitemOpen
  \bibfield  {author} {\bibinfo {author} {\bibfnamefont {S.}~\bibnamefont
  {Peredkov}}, \bibinfo {author} {\bibfnamefont {M.}~\bibnamefont {Neeb}},
  \bibinfo {author} {\bibfnamefont {W.}~\bibnamefont {Eberhardt}}, \bibinfo
  {author} {\bibfnamefont {J.}~\bibnamefont {Meyer}}, \bibinfo {author}
  {\bibfnamefont {M.}~\bibnamefont {Tombers}}, \bibinfo {author} {\bibfnamefont
  {H.}~\bibnamefont {Kampschulte}}, \ and\ \bibinfo {author} {\bibfnamefont
  {G.}~\bibnamefont {Niedner-Schatteburg}},\ }\href@noop {} {\bibfield
  {journal} {\bibinfo  {journal} {Phys. Rev. Lett.},\ }\textbf {\bibinfo
  {volume} {107}},\ \bibinfo {pages} {233401} (\bibinfo {year}
  {2011})}\BibitemShut {NoStop}%
\bibitem [{\citenamefont {Niemeyer}\ \emph {et~al.}(2012)\citenamefont
  {Niemeyer}, \citenamefont {Hirsch}, \citenamefont {Zamudio-Bayer},
  \citenamefont {Langenberg}, \citenamefont {Vogel}, \citenamefont {Kossick},
  \citenamefont {Ebrecht}, \citenamefont {Egashira}, \citenamefont {Terasaki},
  \citenamefont {M{\"o}ller}, \citenamefont {v.~Issendorff},\ and\
  \citenamefont {Lau}}]{Niemeyer12}%
  \BibitemOpen
  \bibfield  {author} {\bibinfo {author} {\bibfnamefont {M.}~\bibnamefont
  {Niemeyer}}, \bibinfo {author} {\bibfnamefont {K.}~\bibnamefont {Hirsch}},
  \bibinfo {author} {\bibfnamefont {V.}~\bibnamefont {Zamudio-Bayer}}, \bibinfo
  {author} {\bibfnamefont {A.}~\bibnamefont {Langenberg}}, \bibinfo {author}
  {\bibfnamefont {M.}~\bibnamefont {Vogel}}, \bibinfo {author} {\bibfnamefont
  {M.}~\bibnamefont {Kossick}}, \bibinfo {author} {\bibfnamefont
  {C.}~\bibnamefont {Ebrecht}}, \bibinfo {author} {\bibfnamefont
  {K.}~\bibnamefont {Egashira}}, \bibinfo {author} {\bibfnamefont
  {A.}~\bibnamefont {Terasaki}}, \bibinfo {author} {\bibfnamefont
  {T.}~\bibnamefont {M{\"o}ller}}, \bibinfo {author} {\bibfnamefont
  {B.}~\bibnamefont {v.~Issendorff}}, \ and\ \bibinfo {author} {\bibfnamefont
  {J.~T.}\ \bibnamefont {Lau}},\ }\href@noop {} {\bibfield  {journal} {\bibinfo
   {journal} {Phys. Rev. Lett.},\ }\textbf {\bibinfo {volume} {108}},\ \bibinfo
  {pages} {057201} (\bibinfo {year} {2012})}\BibitemShut {NoStop}%
\bibitem [{\citenamefont {Vogel}\ \emph {et~al.}(2012)\citenamefont {Vogel},
  \citenamefont {Kasigkeit}, \citenamefont {Hirsch}, \citenamefont
  {Langenberg}, \citenamefont {Rittmann}, \citenamefont {Zamudio-Bayer},
  \citenamefont {Kulesza}, \citenamefont {Mitric}, \citenamefont {M{\"o}ller},
  \citenamefont {von Issendorff},\ and\ \citenamefont {Lau}}]{Vogel12}%
  \BibitemOpen
  \bibfield  {author} {\bibinfo {author} {\bibfnamefont {M.}~\bibnamefont
  {Vogel}}, \bibinfo {author} {\bibfnamefont {C.}~\bibnamefont {Kasigkeit}},
  \bibinfo {author} {\bibfnamefont {K.}~\bibnamefont {Hirsch}}, \bibinfo
  {author} {\bibfnamefont {A.}~\bibnamefont {Langenberg}}, \bibinfo {author}
  {\bibfnamefont {J.}~\bibnamefont {Rittmann}}, \bibinfo {author}
  {\bibfnamefont {V.}~\bibnamefont {Zamudio-Bayer}}, \bibinfo {author}
  {\bibfnamefont {A.}~\bibnamefont {Kulesza}}, \bibinfo {author} {\bibfnamefont
  {R.}~\bibnamefont {Mitric}}, \bibinfo {author} {\bibfnamefont
  {T.}~\bibnamefont {M{\"o}ller}}, \bibinfo {author} {\bibfnamefont
  {B.}~\bibnamefont {von Issendorff}}, \ and\ \bibinfo {author} {\bibfnamefont
  {J.~T.}\ \bibnamefont {Lau}},\ }\href@noop {} {\bibfield  {journal} {\bibinfo
   {journal} {Phys. Rev. B},\ }\textbf {\bibinfo {volume} {85}},\ \bibinfo
  {pages} {195454} (\bibinfo {year} {2012})}\BibitemShut {NoStop}%
\bibitem [{\citenamefont {Hirsch}\ \emph {et~al.}(2009)\citenamefont {Hirsch},
  \citenamefont {Lau}, \citenamefont {Klar}, \citenamefont {Langenberg},
  \citenamefont {Probst}, \citenamefont {Rittmann}, \citenamefont {Vogel},
  \citenamefont {Zamudio-Bayer}, \citenamefont {M{\"o}ller},\ and\
  \citenamefont {Issendorff}}]{Hirsch09}%
  \BibitemOpen
  \bibfield  {author} {\bibinfo {author} {\bibfnamefont {K.}~\bibnamefont
  {Hirsch}}, \bibinfo {author} {\bibfnamefont {J.~T.}\ \bibnamefont {Lau}},
  \bibinfo {author} {\bibfnamefont {P.}~\bibnamefont {Klar}}, \bibinfo {author}
  {\bibfnamefont {A.}~\bibnamefont {Langenberg}}, \bibinfo {author}
  {\bibfnamefont {J.}~\bibnamefont {Probst}}, \bibinfo {author} {\bibfnamefont
  {J.}~\bibnamefont {Rittmann}}, \bibinfo {author} {\bibfnamefont
  {M.}~\bibnamefont {Vogel}}, \bibinfo {author} {\bibfnamefont
  {V.}~\bibnamefont {Zamudio-Bayer}}, \bibinfo {author} {\bibfnamefont
  {T.}~\bibnamefont {M{\"o}ller}}, \ and\ \bibinfo {author} {\bibfnamefont
  {B.~v.}\ \bibnamefont {Issendorff}},\ }\href@noop {} {\bibfield  {journal}
  {\bibinfo  {journal} {J. Phys. B},\ }\textbf {\bibinfo {volume} {42}},\
  \bibinfo {pages} {154029} (\bibinfo {year} {2009})}\BibitemShut {NoStop}%
\bibitem [{\citenamefont {Thole}\ \emph {et~al.}(1992)\citenamefont {Thole},
  \citenamefont {Carra}, \citenamefont {Sette},\ and\ \citenamefont {van~der
  Laan}}]{Thole92}%
  \BibitemOpen
  \bibfield  {author} {\bibinfo {author} {\bibfnamefont {B.~T.}\ \bibnamefont
  {Thole}}, \bibinfo {author} {\bibfnamefont {P.}~\bibnamefont {Carra}},
  \bibinfo {author} {\bibfnamefont {F.}~\bibnamefont {Sette}}, \ and\ \bibinfo
  {author} {\bibfnamefont {G.}~\bibnamefont {van~der Laan}},\ }\Doi
  {10.1103/PhysRevLett.68.1943} {\bibfield  {journal} {\bibinfo  {journal}
  {Phys. Rev. Lett.},\ }\textbf {\bibinfo {volume} {68}},\ \bibinfo {pages}
  {1943} (\bibinfo {year} {1992})}\BibitemShut {NoStop}%
\bibitem [{\citenamefont {Carra}\ \emph {et~al.}(1993)\citenamefont {Carra},
  \citenamefont {Thole}, \citenamefont {Altarelli},\ and\ \citenamefont
  {Wang}}]{Carra93}%
  \BibitemOpen
  \bibfield  {author} {\bibinfo {author} {\bibfnamefont {P.}~\bibnamefont
  {Carra}}, \bibinfo {author} {\bibfnamefont {B.~T.}\ \bibnamefont {Thole}},
  \bibinfo {author} {\bibfnamefont {M.}~\bibnamefont {Altarelli}}, \ and\
  \bibinfo {author} {\bibfnamefont {X.}~\bibnamefont {Wang}},\ }\Doi
  {10.1103/PhysRevLett.70.694} {\bibfield  {journal} {\bibinfo  {journal}
  {Phys. Rev. Lett.},\ }\textbf {\bibinfo {volume} {70}},\ \bibinfo {pages}
  {694} (\bibinfo {year} {1993})}\BibitemShut {NoStop}%
\bibitem [{\citenamefont {Ohara}\ \emph {et~al.}(2003)\citenamefont {Ohara},
  \citenamefont {Koyasu}, \citenamefont {Nakajima},\ and\ \citenamefont
  {Kaya}}]{Ohara03}%
  \BibitemOpen
  \bibfield  {author} {\bibinfo {author} {\bibfnamefont {M.}~\bibnamefont
  {Ohara}}, \bibinfo {author} {\bibfnamefont {K.}~\bibnamefont {Koyasu}},
  \bibinfo {author} {\bibfnamefont {A.}~\bibnamefont {Nakajima}}, \ and\
  \bibinfo {author} {\bibfnamefont {K.}~\bibnamefont {Kaya}},\ }\href@noop {}
  {\bibfield  {journal} {\bibinfo  {journal} {Chem. Phys. Lett.},\ }\textbf
  {\bibinfo {volume} {371}},\ \bibinfo {pages} {490} (\bibinfo {year}
  {2003})}\BibitemShut {NoStop}%
\bibitem [{\citenamefont {Janssens}\ \emph {et~al.}(2007)\citenamefont
  {Janssens}, \citenamefont {Gruene}, \citenamefont {Meijer}, \citenamefont
  {Woste}, \citenamefont {Lievens},\ and\ \citenamefont
  {Fielicke}}]{Janssens07}%
  \BibitemOpen
  \bibfield  {author} {\bibinfo {author} {\bibfnamefont {E.}~\bibnamefont
  {Janssens}}, \bibinfo {author} {\bibfnamefont {P.}~\bibnamefont {Gruene}},
  \bibinfo {author} {\bibfnamefont {G.}~\bibnamefont {Meijer}}, \bibinfo
  {author} {\bibfnamefont {L.}~\bibnamefont {Woste}}, \bibinfo {author}
  {\bibfnamefont {P.}~\bibnamefont {Lievens}}, \ and\ \bibinfo {author}
  {\bibfnamefont {A.}~\bibnamefont {Fielicke}},\ }\href@noop {} {\bibfield
  {journal} {\bibinfo  {journal} {Phys. Rev. Lett.},\ }\textbf {\bibinfo
  {volume} {99}},\ \bibinfo {pages} {063401} (\bibinfo {year}
  {2007})}\BibitemShut {NoStop}%
\bibitem [{\citenamefont {Kirkpatrick}\ \emph {et~al.}(1983)\citenamefont
  {Kirkpatrick}, \citenamefont {Gelatt},\ and\ \citenamefont
  {Vecchi}}]{Kirkpatrick1983}%
  \BibitemOpen
  \bibfield  {author} {\bibinfo {author} {\bibfnamefont {S.}~\bibnamefont
  {Kirkpatrick}}, \bibinfo {author} {\bibfnamefont {C.~D.}\ \bibnamefont
  {Gelatt}}, \ and\ \bibinfo {author} {\bibfnamefont {M.~P.}\ \bibnamefont
  {Vecchi}},\ }\href@noop {} {\bibfield  {journal} {\bibinfo  {journal}
  {Science},\ }\textbf {\bibinfo {volume} {220}},\ \bibinfo {pages} {671}
  (\bibinfo {year} {1983})}\BibitemShut {NoStop}%
\bibitem [{\citenamefont {Jackson}\ \emph {et~al.}(2004)\citenamefont
  {Jackson}, \citenamefont {Horoi}, \citenamefont {Chaudhuri}, \citenamefont
  {Frauenheim},\ and\ \citenamefont {Shvartsburg}}]{Jackson04}%
  \BibitemOpen
  \bibfield  {author} {\bibinfo {author} {\bibfnamefont {K.~A.}\ \bibnamefont
  {Jackson}}, \bibinfo {author} {\bibfnamefont {M.}~\bibnamefont {Horoi}},
  \bibinfo {author} {\bibfnamefont {I.}~\bibnamefont {Chaudhuri}}, \bibinfo
  {author} {\bibfnamefont {T.}~\bibnamefont {Frauenheim}}, \ and\ \bibinfo
  {author} {\bibfnamefont {A.~A.}\ \bibnamefont {Shvartsburg}},\ }\href@noop {}
  {\bibfield  {journal} {\bibinfo  {journal} {Phys. Rev. Lett.},\ }\textbf
  {\bibinfo {volume} {93}},\ \bibinfo {pages} {013401} (\bibinfo {year}
  {2004})}\BibitemShut {NoStop}%
\bibitem [{SM_()}]{SM_MnSi}%
  \BibitemOpen
  \href@noop {} {}\bibinfo {note} {See Supplemental Material at [URL] for
  details of the global optimization procedure and the calculation of the
  self-interaction error, as well as simulated harmonic vibrational spectra and
  xyz-coordinates of the $\mathrm{MnSi}_n^+$ ground state
  structures}\BibitemShut {NoStop}%
\bibitem [{\citenamefont {Adamo}\ and\ \citenamefont {Barone}(1999)}]{Adamo99}%
  \BibitemOpen
  \bibfield  {author} {\bibinfo {author} {\bibfnamefont {C.}~\bibnamefont
  {Adamo}}\ and\ \bibinfo {author} {\bibfnamefont {V.}~\bibnamefont {Barone}},\
  }\href@noop {} {\bibfield  {journal} {\bibinfo  {journal} {J. Chem. Phys.},\
  }\textbf {\bibinfo {volume} {110}},\ \bibinfo {pages} {6158} (\bibinfo {year}
  {1999})}\BibitemShut {NoStop}%
\bibitem [{TUR()}]{TURBOMOLE3}%
  \BibitemOpen
  \href@noop {} {}\bibinfo {note} {{TURBOMOLE} V6.0 2009}\BibitemShut {NoStop}%
\bibitem [{\citenamefont {Perdew}\ and\ \citenamefont
  {Zunger}(1981)}]{Perdew81}%
  \BibitemOpen
  \bibfield  {author} {\bibinfo {author} {\bibfnamefont {J.~P.}\ \bibnamefont
  {Perdew}}\ and\ \bibinfo {author} {\bibfnamefont {A.}~\bibnamefont
  {Zunger}},\ }\Doi {10.1103/PhysRevB.23.5048} {\bibfield  {journal} {\bibinfo
  {journal} {Phys. Rev. B},\ }\textbf {\bibinfo {volume} {23}},\ \bibinfo
  {pages} {5048} (\bibinfo {year} {1981})}\BibitemShut {NoStop}%
\bibitem [{\citenamefont {K\"orzd\"orfer}\ \emph {et~al.}(2009)\citenamefont
  {K\"orzd\"orfer}, \citenamefont {K\"ummel}, \citenamefont {Marom},\ and\
  \citenamefont {Kronik}}]{Korzdorfer09}%
  \BibitemOpen
  \bibfield  {author} {\bibinfo {author} {\bibfnamefont {T.}~\bibnamefont
  {K\"orzd\"orfer}}, \bibinfo {author} {\bibfnamefont {S.}~\bibnamefont
  {K\"ummel}}, \bibinfo {author} {\bibfnamefont {N.}~\bibnamefont {Marom}}, \
  and\ \bibinfo {author} {\bibfnamefont {L.}~\bibnamefont {Kronik}},\
  }\href@noop {} {\bibfield  {journal} {\bibinfo  {journal} {Phys. Rev. B},\
  }\textbf {\bibinfo {volume} {79}},\ \bibinfo {pages} {201205(R)} (\bibinfo
  {year} {2009})}\BibitemShut {NoStop}%
\bibitem [{\citenamefont {K\"orzd\"orfer}\ \emph {et~al.}(2010)\citenamefont
  {K\"orzd\"orfer}, \citenamefont {K\"ummel}, \citenamefont {Marom},\ and\
  \citenamefont {Kronik}}]{Korzdorfer10b}%
  \BibitemOpen
  \bibfield  {author} {\bibinfo {author} {\bibfnamefont {T.}~\bibnamefont
  {K\"orzd\"orfer}}, \bibinfo {author} {\bibfnamefont {S.}~\bibnamefont
  {K\"ummel}}, \bibinfo {author} {\bibfnamefont {N.}~\bibnamefont {Marom}}, \
  and\ \bibinfo {author} {\bibfnamefont {L.}~\bibnamefont {Kronik}},\ }\Doi
  {10.1103/PhysRevB.82.129903} {\bibfield  {journal} {\bibinfo  {journal}
  {Phys. Rev B},\ }\textbf {\bibinfo {volume} {82}},\ \bibinfo {pages} {129903}
  (\bibinfo {year} {2010})}\BibitemShut {NoStop}%
\bibitem [{\citenamefont {Lyon}\ \emph {et~al.}(2009)\citenamefont {Lyon},
  \citenamefont {Gruene}, \citenamefont {Fielicke}, \citenamefont {Meijer},
  \citenamefont {Janssens}, \citenamefont {Claes},\ and\ \citenamefont
  {Lievens}}]{Lyon09}%
  \BibitemOpen
  \bibfield  {author} {\bibinfo {author} {\bibfnamefont {J.~T.}\ \bibnamefont
  {Lyon}}, \bibinfo {author} {\bibfnamefont {P.}~\bibnamefont {Gruene}},
  \bibinfo {author} {\bibfnamefont {A.}~\bibnamefont {Fielicke}}, \bibinfo
  {author} {\bibfnamefont {G.}~\bibnamefont {Meijer}}, \bibinfo {author}
  {\bibfnamefont {E.}~\bibnamefont {Janssens}}, \bibinfo {author}
  {\bibfnamefont {P.}~\bibnamefont {Claes}}, \ and\ \bibinfo {author}
  {\bibfnamefont {P.}~\bibnamefont {Lievens}},\ }\href@noop {} {\bibfield
  {journal} {\bibinfo  {journal} {J. Am. Chem. Soc.},\ }\textbf {\bibinfo
  {volume} {131}},\ \bibinfo {pages} {1115} (\bibinfo {year}
  {2009})}\BibitemShut {NoStop}%
\bibitem [{\citenamefont {Lau}\ \emph {et~al.}(2011)\citenamefont {Lau},
  \citenamefont {Vogel}, \citenamefont {Langenberg}, \citenamefont {Hirsch},
  \citenamefont {Rittmann}, \citenamefont {Zamudio-Bayer}, \citenamefont
  {M{\"o}ller},\ and\ \citenamefont {von Issendorff}}]{Lau11}%
  \BibitemOpen
  \bibfield  {author} {\bibinfo {author} {\bibfnamefont {J.~T.}\ \bibnamefont
  {Lau}}, \bibinfo {author} {\bibfnamefont {M.}~\bibnamefont {Vogel}}, \bibinfo
  {author} {\bibfnamefont {A.}~\bibnamefont {Langenberg}}, \bibinfo {author}
  {\bibfnamefont {K.}~\bibnamefont {Hirsch}}, \bibinfo {author} {\bibfnamefont
  {J.}~\bibnamefont {Rittmann}}, \bibinfo {author} {\bibfnamefont
  {V.}~\bibnamefont {Zamudio-Bayer}}, \bibinfo {author} {\bibfnamefont
  {T.}~\bibnamefont {M{\"o}ller}}, \ and\ \bibinfo {author} {\bibfnamefont
  {B.}~\bibnamefont {von Issendorff}},\ }\href@noop {} {\bibfield  {journal}
  {\bibinfo  {journal} {J. Chem. Phys.},\ }\textbf {\bibinfo {volume} {134}},\
  \bibinfo {pages} {041102} (\bibinfo {year} {2011})}\BibitemShut {NoStop}%
\bibitem [{\citenamefont {Kronik}\ \emph {et~al.}(2006)\citenamefont {Kronik},
  \citenamefont {Makmal}, \citenamefont {Tiago}, \citenamefont {Alemany},
  \citenamefont {Jain}, \citenamefont {Huang}, \citenamefont {Saad},\ and\
  \citenamefont {Chelikowsky}}]{Kronik2006}%
  \BibitemOpen
  \bibfield  {author} {\bibinfo {author} {\bibfnamefont {L.}~\bibnamefont
  {Kronik}}, \bibinfo {author} {\bibfnamefont {A.}~\bibnamefont {Makmal}},
  \bibinfo {author} {\bibfnamefont {M.~L.}\ \bibnamefont {Tiago}}, \bibinfo
  {author} {\bibfnamefont {M.~M.~G.}\ \bibnamefont {Alemany}}, \bibinfo
  {author} {\bibfnamefont {M.}~\bibnamefont {Jain}}, \bibinfo {author}
  {\bibfnamefont {X.}~\bibnamefont {Huang}}, \bibinfo {author} {\bibfnamefont
  {Y.}~\bibnamefont {Saad}}, \ and\ \bibinfo {author} {\bibfnamefont {J.~R.}\
  \bibnamefont {Chelikowsky}},\ }\href@noop {} {\bibfield  {journal} {\bibinfo
  {journal} {Phys. Status Solidi B},\ }\textbf {\bibinfo {volume} {243}},\
  \bibinfo {pages} {1063} (\bibinfo {year} {2006})}\BibitemShut {NoStop}%
\bibitem [{\citenamefont {Perdew}\ \emph {et~al.}(1996)\citenamefont {Perdew},
  \citenamefont {Burke},\ and\ \citenamefont {Ernzerhof}}]{Perdew1996}%
  \BibitemOpen
  \bibfield  {author} {\bibinfo {author} {\bibfnamefont {J.~P.}\ \bibnamefont
  {Perdew}}, \bibinfo {author} {\bibfnamefont {K.}~\bibnamefont {Burke}}, \
  and\ \bibinfo {author} {\bibfnamefont {M.}~\bibnamefont {Ernzerhof}},\
  }\href@noop {} {\bibfield  {journal} {\bibinfo  {journal} {Phys. Rev.
  Lett.},\ }\textbf {\bibinfo {volume} {77}},\ \bibinfo {pages} {3865}
  (\bibinfo {year} {1996})}\BibitemShut {NoStop}%
\bibitem [{\citenamefont {K{\"o}rzd{\"o}rfer}\ and\ \citenamefont
  {K{\"u}mmel}(2010)}]{Korzdorfer10a}%
  \BibitemOpen
  \bibfield  {author} {\bibinfo {author} {\bibfnamefont {T.}~\bibnamefont
  {K{\"o}rzd{\"o}rfer}}\ and\ \bibinfo {author} {\bibfnamefont
  {S.}~\bibnamefont {K{\"u}mmel}},\ }\href@noop {} {\bibfield  {journal}
  {\bibinfo  {journal} {Phys. Rev B},\ }\textbf {\bibinfo {volume} {82}},\
  \bibinfo {pages} {155206} (\bibinfo {year} {2010})}\BibitemShut {NoStop}%
\bibitem [{\citenamefont {Wu}\ \emph {et~al.}(2004)\citenamefont {Wu},
  \citenamefont {Hortamani}, \citenamefont {Kratzer},\ and\ \citenamefont
  {Scheffler}}]{Wu04a}%
  \BibitemOpen
  \bibfield  {author} {\bibinfo {author} {\bibfnamefont {H.}~\bibnamefont
  {Wu}}, \bibinfo {author} {\bibfnamefont {M.}~\bibnamefont {Hortamani}},
  \bibinfo {author} {\bibfnamefont {P.}~\bibnamefont {Kratzer}}, \ and\
  \bibinfo {author} {\bibfnamefont {M.}~\bibnamefont {Scheffler}},\ }\href
  {http://link.aps.org/doi/10.1103/PhysRevLett.92.237202} {\bibfield  {journal}
  {\bibinfo  {journal} {Phys. Rev. Lett.},\ }\textbf {\bibinfo {volume} {92}},\
  \bibinfo {pages} {237202} (\bibinfo {year} {2004})}\BibitemShut {NoStop}%
\end{thebibliography}
\end{document}